\documentclass[final,5p,times,twocolumn]{elsarticle}

\usepackage{amssymb}
\journal{Physica E}

\begin{document}

\begin{frontmatter}



\title{Effects of noise on synchronization phenomena exhibited by mean-field coupled limit cycle oscillators with two natural frequencies}


\author[label1]{Keiji Okumura}
\author[label1]{Akihisa Ichiki}
\author[label1]{Masatoshi Shiino}

\address[label1]{Department of Physics, Faculty of Science, Tokyo Institute of Technology, 2-12-1 Oh-okayama, Meguro-ku, Tokyo 152-8551, Japan}

\begin{abstract}
Relationships between inter-cluster synchronization phenomena and external noise are studied on the basis of noise level-free analysis. We consider a mean-field model of ensembles of coupled limit cycle oscillators with two natural frequencies, which are subjected to external white Gaussian noise. Using a nonlinear Fokker-Planck equation approach, we \textit{analytically} derive the order parameter equations associated with nonequilibrium phase transitions in the thermodynamic limit. Solving them numerically, we systematically investigate the model parameter dependences of the appearance and disappearance of synchronization phenomena. Demonstrating bifurcations from chaotic attractors in the deterministic limit to limit cycle attractors with increasing noise intensity, we confirm the occurrence of nonequilibrium phase transitions including inter-cluster synchronization induced by external white Gaussian noise.
\end{abstract}

\begin{keyword}
Noise induced synchronization \sep Mean-field model \sep Nonlinear Fokker-Planck equation \sep Nonequilibrium phase transitions \sep Stochastic limit cycle oscillators
\end{keyword}

\end{frontmatter}

% \linenumbers

\section{Introduction}
\label{sec:intro}
Effects of noise on synchronization phenomena in oscillatory systems have recently attracted much attention from many researchers. Synchronization phenomena are ubiquitous ones observed in various fields of natural sciences \cite{Pikovsky01TEXT}. In neurosciences, neurons in the basal ganglia exhibit more synchrony in Parkinson's disease than in normal state, suggesting that neural information coding is closely related to the synchronization phenomena \cite{Hammond07}. For such a reason, to study how the noise exerts its influence on the structure of synchronization will be of paramount importance from the viewpoint of nonlinear dynamical control involving changes in synchrony of oscillatory systems.

On one hand, someone would intuitively suppose that the presence of noise might deteriorate the degree of synchronization of oscillatory systems. Breakdown of synchronization due to external noise of an ensemble of limit cycle oscillators was reported \cite{Shiino85}. On the other hand, the opposite phenomena of noise-induced synchronization are becoming an active field of the study of nonlinear dynamical systems \cite{Kurrer95,Kanamaru03,Wang00,Teramae04,Kawamura08,Hakim94}. Noise induced synchronization in coupled excitable systems/active rotators is investigated both analytically \cite{Kurrer95,Kanamaru03} and numerically \cite{Wang00}. Among analytical studies on synchronization phenomena of ensemble of limit cycle oscillators is the phase reduction analysis. Such type of studies revealed the effects of common noise on synchronization of uncoupled oscillators \cite{Teramae04} and uncoupled two populations of oscillators \cite{Kawamura08}. These studies, however, are restricted to the case with weak noise. Numerical approaches to noise induced synchronization of coupled limit cycle oscillators might include the possibilities that what are so obtained are subjected to a finite size effect without taking the thermodynamic limit as in Ref. \cite{Hakim94}.

To avoid such limitations, one may take nonlinear Fokker-Planck equation approaches, which are very closely related to the study of (thermal) equilibrium phase transitions. Presenting the validity of an H theorem for a nonlinear Fokker-Planck equation for a stochastic system of mean-field coupled overdamped oscillators, Shiino studied statistical behaviors of the system exhibiting equilibrium phase transitions \cite{Shiino87} (see also Ref. \cite{Desai78,Dawson83}). Furthermore, in the case of nonequilibrium phase transitions, one can take advantage of using nonlinear Fokker-Planck equations to exactly derive the time evolution of the order-parameter equations in the thermodynamic limit for systems of nonlinearly mean-field coupled oscillators. There, the nonlinear dynamic aspects of phase transitions were investigated, yielding the occurrence of chaos-nonchaos phase transitions and those including intra-cluster synchronization induced by external noise \cite{Shiino01}. Purely noise induced phase transitions involving chaos-nonchaos phase transitions in an ensemble of limit cycle oscillators were also explored \cite{Ichiki07}. A nonlinearly coupled system with time delay was found to exhibit various types of nonequilibrium phase transitions \cite{Shiino07}.

In the present study, we apply the nonlinear Fokker-Planck approach to noise induced inter-cluster synchronization phenomena of coupled limit cycle oscillators. Dealing with a solvable model based on the mean-field concept to derive order parameter equations \cite{Shiino01,Ichiki07,Shiino07}, we study the effects of noise on synchronization in the thermodynamic limit and nonequilibrium stationary states, for which numerical approaches to solving stochastic differential equations do not properly work. Our results show the appearance of nonequilibrium phase transitions involving inter-cluster synchronization. Since the model of oscillators is based on that of analog neural networks, behaviors of noise-induced synchronization in the system will be of practical interest.

\section{Model and nonlinear Fokker-Planck equation approach}
\label{sec:model}
We consider a system of coupled limit cycle oscillators consisting of two subsystems. The oscillators are coupled via nonlinear global interactions and subjected to independent additive and multiplicative external noise. The model dynamics we are concerned with is described by a set of Langevin equations \cite{Ichiki07}:\\
for $i \in N_{1}=\{1,\cdots,n_{1}\}$ , 
\begin{eqnarray}
\lefteqn{ \frac{\textrm{d}z_{1i}^{(\mu)}}{\textrm{d}t} = -a_{1}^{(\mu)}z_{1i}^{(\mu)}+\displaystyle \sum_{j\in N_{1}}J_{11ij}^{(\mu)}F_{1}^{(\mu)}(b_{1}^{(\mu,x)}z_{1j}^{(x)}+b_{1}^{(\mu,y)}z_{1j}^{(y)})}\label{eq:model1}\nonumber\\
&&\quad+\displaystyle \epsilon\sum_{k\in N_{2}}J_{21ik}^{(\mu)}F_{2}^{(\mu)}(b_{2}^{(\mu,x)}z_{2k}^{(x)}+b_{2}^{(\mu,y)}z_{2k}^{(y)})+\eta_{1i}^{(\mu)}(t),
\end{eqnarray}
for $i \in N_{2}=\{n_{1}+1,\cdots,n_{1}+n_{2}\}$ , 
\begin{eqnarray}
\lefteqn{ \frac{\textrm{d}z_{2i}^{(\mu)}}{\textrm{d}t} = -a_{2}^{(\mu)}z_{2i}^{(\mu)}+\displaystyle \sum_{j\in N_{2}}J_{22ij}^{(\mu)}F_{2}^{(\mu)}(b_{2}^{(\mu,x)}z_{2j}^{(x)}+b_{2}^{(\mu,y)}z_{2j}^{(y)})}\label{eq:model2}\nonumber\\
&&\quad+\displaystyle \epsilon\sum_{k\in N_{1}}J_{12ik}^{(\mu)}F_{1}^{(\mu)}(b_{1}^{(\mu,x)}z_{1k}^{(x)}+b_{1}^{(\mu,y)}z_{1k}^{(y)})+\eta_{2i}^{(\mu)}(t),
\end{eqnarray}
where $z_{\alpha i}^{(\mu)} \,(\mu=x,y) \, (\alpha=1,2\,(3=1))$ are the 2D-oscillators at site $i$, $a_{\alpha}^{(\mu)},\,b_{\alpha}^{(\mu,\nu)}$ are constants and $F_{\alpha}^{(\mu)}(\cdot)$ are bounded functions specifying the nonlinear couplings. The mean-field coupling strengths $J_{\alpha \beta ij}^{(\mu)}(t)$ are given by
\begin{equation}
J_{\alpha\beta ij}^{(\mu)}(t) = \displaystyle \frac{J_{\alpha\beta}^{(\mu)}}{n_{\alpha}}+\xi_{\alpha \beta ij}^{(\mu)}(t)\label{eq:Coupling}.
\end{equation}
In cluster $1$ and $2$, we postulate the different constraint parameters as
$a_{2}^{(\mu)}=a_{1}^{(\mu)}+\delta^{(\mu)},$
with $\delta^{(\mu)}$ being essentially responsible for the difference of natural frequencies between the oscillators in the two subsystems. The inter-cluster coupling strengths are controlled by $\epsilon$. This parameter takes any real constant value, without constraining to weak connections. The external noise $\eta_{\alpha i}^{(\mu)}(t),\,\xi_{\alpha \beta ij}^{(\mu)}(t)$ are of white Gaussian type,
\begin{eqnarray*}
&&\langle\eta_{\alpha i}^{(\mu)}(t)\rangle=0\label{eq:eta_mean},\\
&&\langle\eta_{\alpha i}^{(\mu)}(t)\eta_{\beta j}^{(\nu)}(t^{\prime})\rangle=2D^{(\mu)}\delta_{ij}\delta_{\mu\nu}\delta_{\alpha\beta}\delta(t-t^{\prime})\label{eq:eta_GaussianWhite},\\
&&\langle\xi_{\alpha \beta ij}^{(\mu)}(t)\rangle=0\label{eq:xi_mean},\\
&&\displaystyle \langle\xi_{\alpha \beta ij}^{(\mu)}(t)\xi_{\gamma \delta kl}^{(\nu)}(t^{\prime})\rangle=\frac{2\tilde{D}^{(\mu)}}{n_{\alpha}}\delta_{ik}\delta_{jl}\delta_{\mu\nu}\delta_{\alpha \gamma}\delta_{\beta \delta}\delta(t-t^{\prime})\label{eq:xi_GaussianWhite}.
\end{eqnarray*}
Note that not common but independent noise is introduced and $\eta_{\alpha i}^{(\mu)}(t),\, \xi_{\alpha \beta ij}^{(\mu)}(t)$ are also independent each other.

In the thermodynamic limit $n_{\alpha} \rightarrow \infty$, influence with Eq. (\ref{eq:Coupling}) on any one oscillator from the others almost surely converges according to the law of large numbers, satisfying the validity of the self-averaging property. Introducing empirical probability density $P(t,z_{1}^{(x)},z_{1}^{(y)},z_{2}^{(x)},z_{2}^{(y)})$, the mean-field coupling terms in Eqs. (\ref{eq:model1}) and (\ref{eq:model2}) are expressed in terms of
\begin{eqnarray}
\lefteqn{ \langle F_{\alpha}^{(\mu)}\rangle\equiv\int \textrm{d}z_{1}^{(x)}\textrm{d}z_{1}^{(y)}\textrm{d}z_{2}^{(x)}\textrm{d}z_{2}^{(y)}F_{\alpha}^{(\mu)}(b_{\alpha}^{(\mu,x)}z_{\alpha}^{(x)}+b_{\alpha}^{(\mu,y)}z_{\alpha}^{(y)}) }\nonumber\\
&&\qquad \qquad \qquad \qquad \qquad \times P(t,z_{1}^{(x)},z_{1}^{(y)},z_{2}^{(x)},z_{2}^{(y)}),\label{eq:MFCoupling1}\\
\lefteqn{ \langle F_{\alpha}^{(\mu)^{2}}\rangle\equiv\int \textrm{d}z_{1}^{(x)}\textrm{d}z_{1}^{(y)}\textrm{d}z_{2}^{(x)}\textrm{d}z_{2}^{(y)}F_{\alpha}^{(\mu)^{2}}(b_{\alpha}^{(\mu,x)}z_{\alpha}^{(x)}+b_{\alpha}^{(\mu,y)}z_{\alpha}^{(y)}) }\nonumber\\
&&\qquad \qquad \qquad \qquad \qquad \times P(t,z_{1}^{(x)},z_{1}^{(y)},z_{2}^{(x)},z_{2}^{(y)}).\label{eq:MFCoupling2}
\end{eqnarray}
The total number of dynamic variables of the system is consequently reduced from $2(n_1+n_2)$ to $4$ in the thermodynamic limit. Then one has a set of the Langevin equations,
\begin{eqnarray}
\lefteqn{ \displaystyle \frac{\textrm{d}z_{\alpha}^{(\mu)}}{\textrm{d}t}=-a_{\alpha}^{(\mu)}z_{\alpha}^{(\mu)}+J_{\alpha\alpha}^{(\mu)}\langle F_{\alpha}^{(\mu)}\rangle+\epsilon J_{\alpha +1,\alpha}^{(\mu)}\langle F_{\alpha +1}^{(\mu)}\rangle+\zeta_{\alpha}^{(\mu)}(t),}\label{eq:effLangevin}
\end{eqnarray}
where the noise is subjected to white Gaussian noise,
\begin{eqnarray*}
&&\langle\zeta_{\alpha}^{(\mu)}(t)\rangle=0,\label{eq:zeta_mean}\\
&&\langle\zeta_{\alpha}^{(\mu)}(t)\zeta_{\beta}^{(\nu)}(t^{\prime})\rangle=2D_{{\tiny \textrm{eff}}\alpha}^{(\mu)}\delta_{\mu\nu}\delta_{\alpha\beta}\delta(t-t^{\prime})\label{eq:zeta_GaussianWhite},\\
&&D_{{\tiny \textrm{eff}}\alpha}^{(\mu)}=D^{(\mu)}+\tilde{D^{(\mu)}}\langle F_{\alpha}^{(\mu)^{2}}\rangle+\epsilon^{2}\tilde{D^{(\mu)}}\langle F_{\alpha +1}^{(\mu)^{2}}\rangle.\nonumber
\end{eqnarray*}
The nonlinear Fokker-Planck equation \cite{Frank05} corresponding to Eqs. (\ref{eq:effLangevin}) reads 
\begin{eqnarray}
\lefteqn{ \frac{\partial}{\partial t}P(t,z_{1}^{(x)},z_{1}^{(y)},z_{2}^{(x)},z_{2}^{(y)})=- \sum_{\alpha=1,2}\sum_{\mu=x,y}\frac{\partial}{\partial z_{\alpha}^{(\mu)}}}\nonumber\\
\lefteqn{ \, \, \left(-a_{\alpha}^{(\mu)}z_{\alpha}^{(\mu)}+J_{\alpha}^{(\mu)}\displaystyle \langle F_{\alpha}^{(\mu)}\rangle+\epsilon J_{\alpha+1}^{(\mu)}\langle F_{\alpha+1}^{(\mu)}\rangle-D_{{\tiny \textrm{eff}}\alpha}^{(\mu)}\frac{\partial}{\partial z_{\alpha}^{(\mu)}}\right)P.}\label{eq:NFPE}
\end{eqnarray}

A Gaussian probability density satisfies Eq. (\ref{eq:NFPE}) as a special solution. Since the H theorem \cite{Shiino01} ensures that the probability density satisfying Eq. (\ref{eq:NFPE}) converges to the Gaussian-form for sufficiently large times, we represent the Gaussian probability density as
\begin{eqnarray*}
\lefteqn{P_{\tiny \textrm{G}}(t,z_{1}^{(x)},z_{1}^{(y)},z_{2}^{(x)},z_{2}^{(y)})=}\nonumber\\
&&\qquad \quad \displaystyle \frac{1}{(2\pi)^{2}\sqrt{\det C_{\tiny \textrm{G}}(t)}}\exp\left[-\frac{1}{2}\textrm{\boldmath $s$}_{\tiny \textrm{G}}^{T}C_{\tiny \textrm{G}}^{-1}(t)\textrm{\boldmath $s$}_{\tiny \textrm{G}}\right],
\end{eqnarray*}
\begin{eqnarray*}
\textrm{\boldmath $s$}_{\tiny \textrm{G}}^{T}&=&(z_{1}^{(x)}-\langle z_{1}^{(x)}\rangle_{\tiny \textrm{G}},\,z_{1}^{(y)}-\langle z_{1}^{(y)}\rangle_{\tiny \textrm{G}},\nonumber\\
&&\qquad \qquad \qquad z_{2}^{(x)}-\langle z_{2}^{(x)}\rangle_{\tiny \textrm{G}},\,z_{2}^{(y)}-\langle z_{2}^{(y)}\rangle_{\tiny \textrm{G}})\nonumber\\
&\equiv&(u_{1}^{(x)},\,u_{1}^{(y)},\,u_{2}^{(x)},\,u_{2}^{(y)}),
\end{eqnarray*}
\begin{eqnarray*}
C_{{\tiny \textrm{G}} ij}(t)=\langle s_{i}s_{j}\rangle_{\tiny \textrm{G}},
\end{eqnarray*}
where $\langle \cdot \rangle_{\tiny \textrm{G}}$ denotes expectation over $P_{\tiny \textrm{G}}$. Now that Eqs. (\ref{eq:MFCoupling1}) and (\ref{eq:MFCoupling2}) are described in terms of the first and second moments, one obtains a set of closed ordinary differential equations involving at most second moments. The time evolution of each moment is calculated from Eq. (\ref{eq:NFPE}) and we write the moment equations as
\begin{eqnarray}
\lefteqn{ \frac{\textrm{d}\langle z_{\alpha}^{(\mu)}\rangle_{\tiny \textrm{G}}}{\textrm{d}t}=-a_{\alpha}^{(\mu)}\langle z_{\alpha}^{(\mu)}\rangle_{\tiny \textrm{G}}+J_{\alpha\alpha}^{(\mu)}\langle F_{\alpha}^{(\mu)}\rangle_{\tiny \textrm{G}}+\epsilon J_{\alpha+1,\alpha}^{(\mu)}\langle F_{\alpha+1}^{(\mu)}\rangle_{\tiny \textrm{G}}}\label{eq:z_mean} \\
\lefteqn{ \frac{\textrm{d}\langle u_{\alpha}^{(\mu)^{2}}\rangle_{\tiny \textrm{G}}}{\textrm{d}t}=-2a_{\alpha}^{(\mu)}\langle u_{\alpha}^{(\mu)^{2}}\rangle_{\tiny \textrm{G}}+2D_{{\tiny \textrm{eff}}\alpha}^{(\mu)}}\label{eq:z_var} \\
\lefteqn{ \frac{\textrm{d}\langle u_{\alpha}^{(\mu)}u_{\beta}^{(\nu)}\rangle_{\tiny \textrm{G}}}{\textrm{d}t}=-(a_{\alpha}^{(\mu)}+a_{\beta}^{(\nu)})\langle u_{\alpha}^{(\mu)}u_{\beta}^{(\nu)}\rangle_{\tiny \textrm{G}},}\label{eq:z_cov} 
\end{eqnarray}
where
\begin{eqnarray*}
(\alpha,\beta,\mu,\nu)&=&(1,1,x,y),\,(2,2,x,y),\,(1,2,x,x),\\
&&(1,2,y,y),\,(1,2,x,y),\,(1,2,y,x).
\end{eqnarray*}
Note that $\langle u_{\alpha}^{(\mu)}u_{\beta}^{(\nu)}\rangle_{\tiny \textrm{G}}\rightarrow 0 \, (t\rightarrow\infty)$, implying that the covariant components of $C_{\tiny \textrm{G}}$ take zero in the stationary states.

For observing qualitative as well as quantitative dynamical behaviors of the system, we proceed to solve numerically the above set of equations. We specify the coupling function as $F_{\alpha}^{(\mu)}(x)=\sin(x)$, for simplicity. Then, $\langle F_{\alpha}^{(\mu)}\rangle_{\tiny \textrm{G}}$ and $\langle F_{\alpha}^{(\mu)^2}\rangle_{\tiny \textrm{G}}$ are calculated as
\begin{eqnarray}
\lefteqn{ \langle F_{\alpha}^{(\mu)}\rangle_{\tiny \textrm{G}} = \sin\left(b_{\alpha}^{(\mu,x)}\langle z_{\alpha}^{(x)}\rangle_{\tiny \textrm{G}}+b_{\alpha}^{(\mu,y)}\langle z_{\alpha}^{(y)}\rangle_{\tiny \textrm{G}}\right)}\nonumber\\
&& \qquad \times\exp\left(-\frac{b_{\alpha}^{(\mu,x)^{2}}}{2}\langle u_{\alpha}^{(x)^{2}}\rangle_{\tiny \textrm{G}}-\frac{b_{\alpha}^{(\mu,y)^{2}}}{2}\langle u_{\alpha}^{(y)^{2}}\rangle_{\tiny \textrm{G}}\right)\label{eq:sin1},\\
\lefteqn{ \langle F_{\alpha}^{(\mu)^{2}}\rangle_{\tiny \textrm{G}} = \displaystyle \frac{1}{2}-\frac{1}{2}\cos\left(2b_{\alpha}^{(\mu,x)}\langle z_{\alpha}^{(x)}\rangle_{\tiny \textrm{G}}+2b_{\alpha}^{(\mu,y)}\langle z_{\alpha}^{(y)}\rangle_{\tiny \textrm{G}}\right)}\nonumber\\
&& \qquad \times\exp\left(-2b_{\alpha}^{(\mu,x)^{2}}\langle u_{\alpha}^{(x)^{2}}\rangle_{\tiny \textrm{G}}-2b_{\alpha}^{(\mu,y)}\langle u_{\alpha}^{(y)^{2}}\rangle_{\tiny \textrm{G}}\right)\label{eq:sin2}.
\end{eqnarray}

\section{Nonequilibrium phase transitions and noise induced inter-cluster synchronization}
The dynamic system of the order parameter equations written by Eqs. (\ref{eq:z_mean})-(\ref{eq:z_cov}) together with Eqs. (\ref{eq:sin1}) and (\ref{eq:sin2}) is nonlinear, allowing various types of bifurcations with changes in the parameters. For the definition of synchronization, let us consider relationships between behaviors of statistical variables in Eqs. (\ref{eq:z_mean})-(\ref{eq:z_cov}) and those of individual oscillators in the system.

First, we address the issue of synchronization phenomena of the coupled oscillators in the stochastic case. On one hand, in deterministic systems of the coupled limit cycle oscillators, the synchronization phenomena are well-defined by the condition that all of the variables of the system undergo a periodic solution with a common time period. On the other hand, in stochastic systems of a finite number of coupled limit cycle oscillators with external noise, each oscillator behaves randomly under the influence of noise. Since the Fokker-Planck equations in this case are linear, the probability densities of the systems, in general, exhibit equilibrium ( {\it i.e.} fixed point type ) probability densities for sufficiently large times. This implies that the order parameters do not oscillate even if individual oscillators behave periodically in the deterministic case. It seems to be difficult to appropriately define the synchronization phenomena in this situation.

To consider the synchronization phenomena in systems of stochastic coupled limit cycle oscillators without the problem mentioned above, we have introduced the concept of taking the thermodynamic limit based on a mean-field model. In general, probability densities of the limit cycle oscillator systems might be multimodal under the influence of weak external noise. To avoid the difficulty of dealing with multimodal probability densities, our model is proposed so that once external noise is introduced to the system, the model constrains the form of the probability densities to the Gaussian ones for sufficiently large times. Note that the probability densities may be allowed to vary in time. In this case, the order parameters of the system may oscillate periodically in time and we can define the synchronization phenomena according to their behaviors. In particular, the oscillations of the mean values have an important meaning from the view point of analogy with the spontaneous magnetizations in spin models and possibly researches in neurosciences.

For these reasons, we consider synchronization phenomena of the oscillators in each cluster $\alpha$, {\it i.e.}, intra-cluster synchronization and between cluster 1 and 2, {\it i.e.}, inter-cluster synchronization. We define intra-cluster synchronization phenomena as (i) the case that ''all of the mean values over the oscillators in a cluster periodically oscillate to form limit cycle attractors, regardless of the behaviors of the variances,'' and inter-cluster synchronization phenomena as (ii) the case that ``(i) occurs as a whole system.'' We focus our attention on the inter-cluster synchronization phenomena.

For simplicity, we only treat the Langevin noise case, i.e., $\tilde{D}^x=\tilde{D}^y=0$, and do not study inter-cluster chaotic synchronization in this paper. From Eqs. (\ref{eq:z_var}), the variances turn out to be constant for sufficiently large times. To investigate the effects of the Langevin noise on the appearance and disappearance of the inter-cluster synchronization phenomena in the thermodynamic limit and nonequilibrium stationary states, we confine ourselves to the bifurcations of limit cycle attractors with changes in the noise intensity. To determine whether an attractor is of limit cycle type or not, we employ two criteria: (a) the largest Lyapunov exponent (LLE) of the attractor estimated numerically nearly equals zero and (b) all of the mean values have a common time period. Note that both of them exclude the case where the attractor is of ergodic torus type.

Numerical calculations were performed with the fourth-order Runge-Kutta method. The model parameter values are $a_{1}^{(x)}=0.5$, $a_{1}^{(y)}=1.0$, $b_{\alpha}^{(x,x)}=0.5$, $b_{\alpha}^{(x,y)}=1.5$, $b_{\alpha}^{(y,x)}=-12.0$, $b_{\alpha}^{(y,y)}=-1.0$, $J_{\alpha\beta}^{(x)}=J_{\beta\alpha}^{(x)}=18.0$, $J_{\alpha\beta}^{(y)}=J_{\beta\alpha}^{(y)}=30.0$, $a_{2}^{(x)}=0.5+\delta^{(x)}$, $a_{2}^{(y)}=1.0+\delta^{(y)}$. Nonequilibrium phase transitions involving the inter-cluster synchronization of the limit cycle oscillators are systematically investigated with changes in the inter-cluster coupling strength $\epsilon$ and difference of constraint parameters $\delta^{(\mu)}$ , and the Langevin noise intensity $D^{(\mu)}$.

\begin{figure}[tb]
  \begin{center} 
  \includegraphics[scale=0.50,angle=-90,clip]{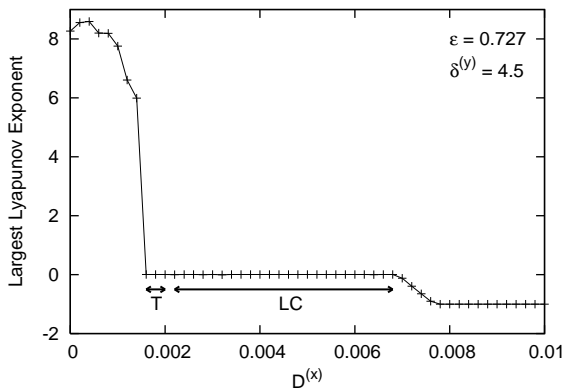}
  \end{center}
  \caption{The largest Lyapunov exponents plotted against the Langevin noise intensity $D^{(x)}$. In the deterministic limit, the LLE has a positive value, implying a chaotic attractor. With an increase of the noise intensity, the LLEs take nearly equal zero, and the attractors changes into limit cycle type via torus type (denoted by LC and T, respectively). We set $D^{(y)}=\delta^{(x)}=0$.}\label{fig:lle}
\end{figure}

To show the appearance of the inter-cluster synchronization due to the effects of noise, we begin with investigating behaviors of the system in the deterministic limit, especially the case where a chaotic attractor appears (Fig. \ref{fig:lle}). With increase of the Langevin noise intensity $D^{(x)}$, the attractor in this case changes into limit cycle type via torus type, suggesting the appearance of the inter-cluster synchronization. Further increase of the noise intensity leads to the negative LLE, implying fixed point type attractors. These results are qualitatively consistent with those observed in Ref. \cite{Hakim94}, which numerically solved a set of Langevin equations to be subjected to the finite size effect.

\begin{figure}[tb]
  \begin{center} 
    \begin{tabular}{c}
      \resizebox{80mm}{!}{\includegraphics[scale=0.50,angle=-90,clip]{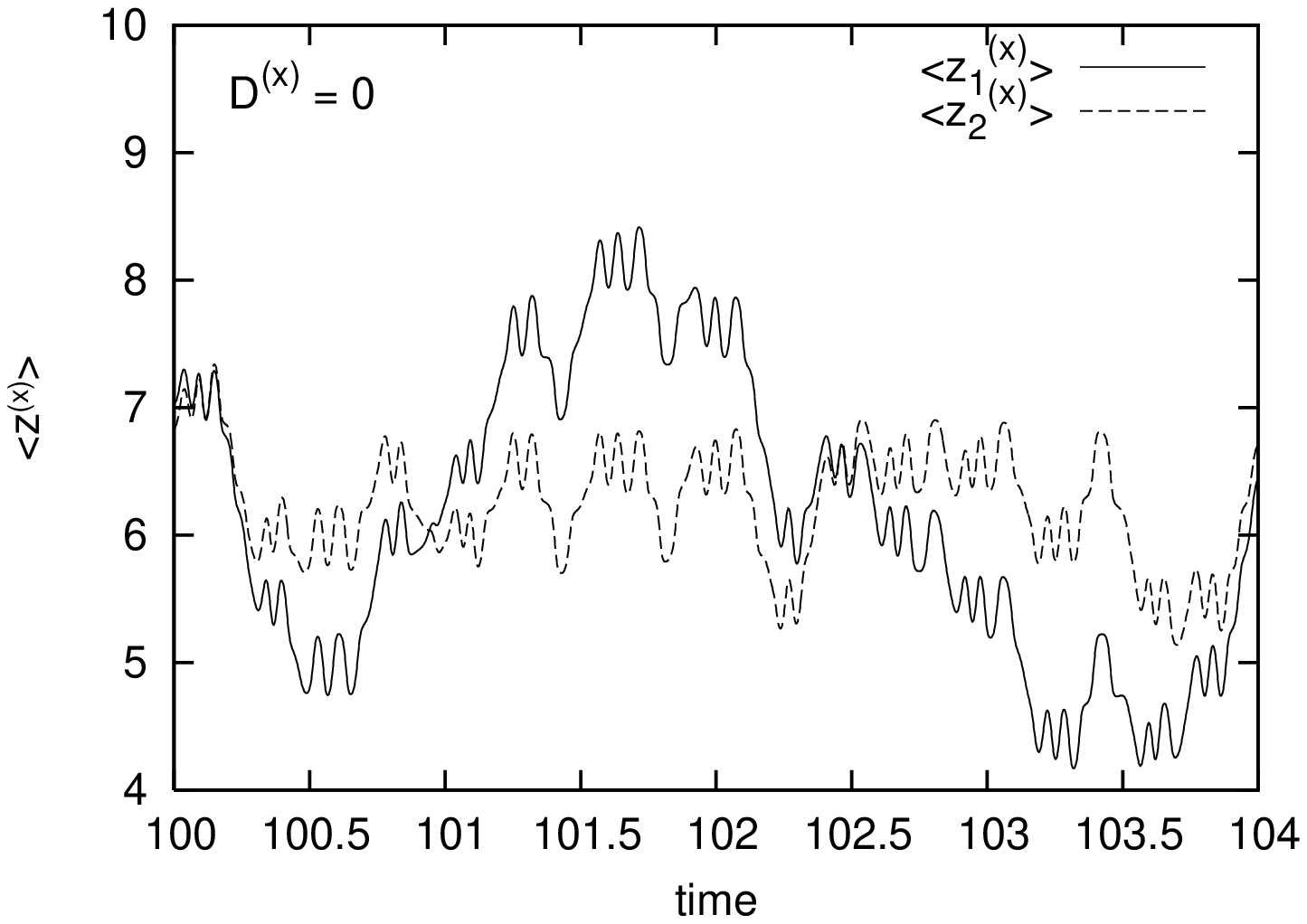}} \\
      \resizebox{80mm}{!}{\includegraphics[scale=0.55,angle=-90,clip]{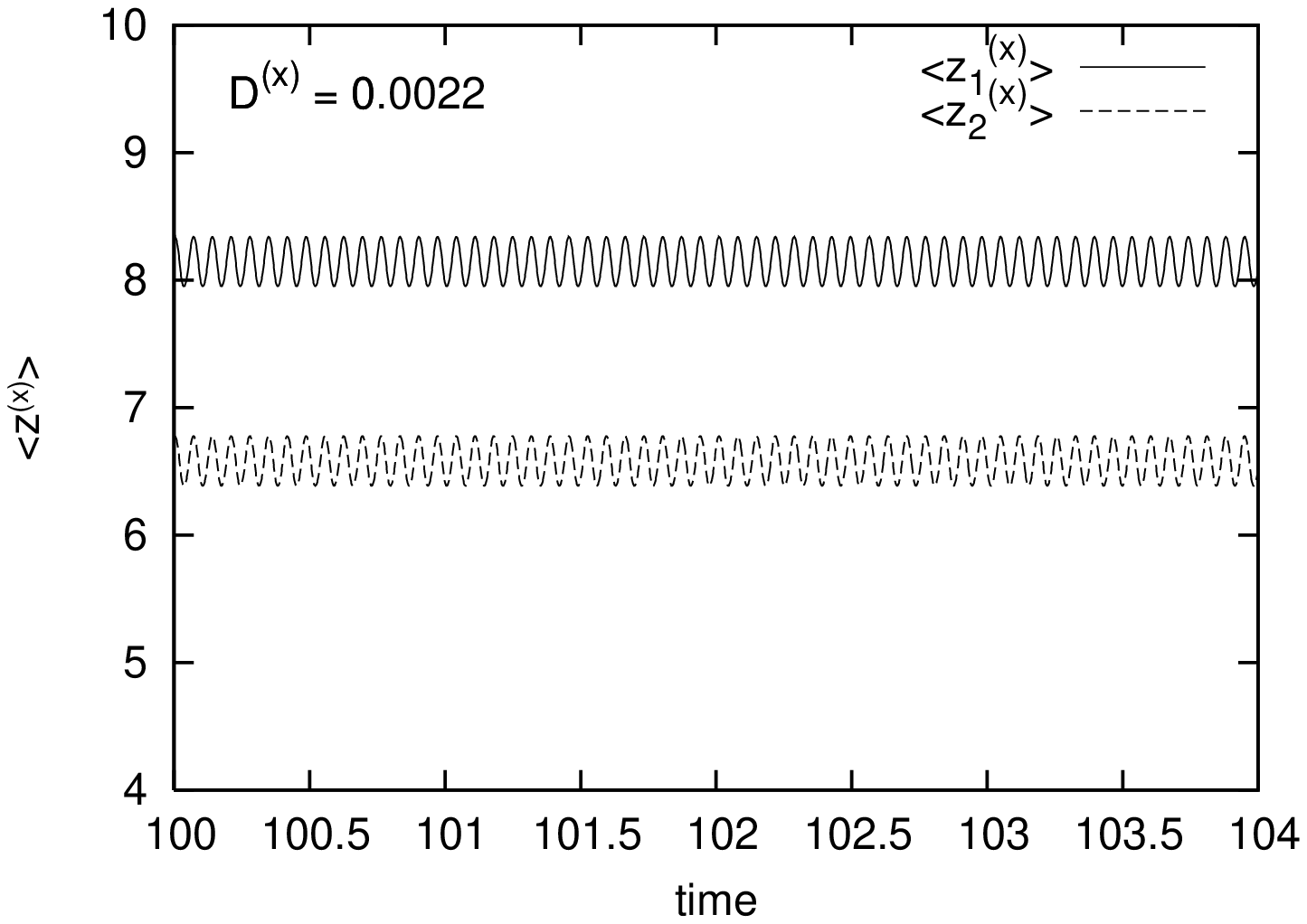}} \\
    \end{tabular}
  \end{center}
 \caption{The time evolution of the mean values $\langle z_{1}^{(x)}\rangle_{\tiny \textrm{G}},\, \langle z_{2}^{(x)}\rangle_{\tiny \textrm{G}}$ of the oscillators of the oscillators in the two clusters, corresponding to the changes from the chaotic to limit cycle attractors in Fig. \ref{fig:lle}. While the oscillators do not behave as synchronous in the deterministic limit (the upper panel), an appropriate level of the Langevin noise intensity induces the inter-cluster synchronization (the lower panel). }\label{fig:TimeEvolution}
\end{figure}

For clearly demonstrating the effects of noise on the inter-cluster synchronization, let us show the temporal evolution of the mean values $\langle z_{1}^{(x)}\rangle_{\tiny \textrm{G}},\, \langle z_{2}^{(x)}\rangle_{\tiny \textrm{G}}$ of the oscillators in cluster 1 and 2 (Fig. \ref{fig:TimeEvolution}). It is easily seen that the strength of noise above a certain level causes the synchronization between the clusters, although both of the oscillators chaotically behave in the deterministic limit.

\section{Concluding remarks}
Using a noise level-free analysis, we have shown the relationships between inter-cluster synchronization phenomena and external noise, especially the occurrence of the inter-cluster synchronization induced by the Langevin noise. Dealing with a mean-field model of ensembles of coupled limit cycle oscillators with two natural frequencies under the influence of external white Gaussian noise, we have analytically derived the order parameter equations in the stationary states, taking advantage of the ingredients of the self-averaging property in the thermodynamic limit together with an H theorem that ensures the convergence of the probability density to the Gaussian-form for sufficiently large times. Solving the order parameter equations numerically, we have systematically investigated the nonequilibrium phase transitions with changes in the inter-cluster coupling strength $\epsilon$ and difference of constraint parameters $\delta^{(\mu)}$, and the Langevin noise intensity $D^{(\mu)}$. The results have shown various interesting bifurcations including the inter-cluster synchronization and chaotic attractors induced by noise. 

Details of the relationship between the parameters involving the inter-cluster coupling strength $\epsilon$ as well as difference of constraint parameters $\delta^{(\mu)}$ and the behaviors of the inter-cluster synchronization, and furthermore the effects of the multiplicative noise on the inter-cluster synchronization will be reported elsewhere.

%\pagebreak

\section*{Acknowledgment}
K.O. acknowledges the financial support from the Global Center of Excellence Program by MEXT, Japan through the ``Nanoscience and Quantum Physics'' Project of the Tokyo Institute of Technology. One of the authors (A.I.) is supported by the Grant-in-Aid for JSPS Fellows No. 20.9513. 



%% References


\bibliographystyle{elsarticle-num}
\bibliography{<your-bib-database>}






% \end{thebibliography}

\end{document}